# Why the same key can open different padlocks?

Fathan Akbar and Mikrajuddin Abdullah[a]

Department of Physics, Bandung Institute of Technology

Jl. Ganesa 10 Bandung 40132, Indonesia

[a]Email: mikrajuddin@gmail.com

**Abstract**

We accidentally observed that one key can open two padlocks from different brands. Based on this observation, here we derived equations to estimate the number of different padlocks that can be produced. We focus on simple padlocks that are usually used in most households. It is made of a rectangular block of metal and crafted at any position. Different keys have different sizes and positions of the crafting (hole) and un-crafting (hill) portions. We limit the study to only padlocks having a maximum of four holes. This is the typical architecture of household padlocks. We observed that the number of different padlocks depends on the minimum size of the holes and hills. We also observe a scaling relationship between the number of different padlocks that can be produced with the size of holes or hills.

## 1. Introduction

Padlocks are widely used worldwide, commonly used for locking fences, warehouses, and similar buildings. Meanwhile for locking houses, key systems that fixed inside the door are commonly opted. The padlock has many advantages, such as it is hardy, less costly compared to other door's lock system, convenient to use, could be carried and used where necessary, and do not need much space [1]. Çelik has discussed the history development of the lock from the Ancient Egypt, Chinese, Iran, Roman, Turk-Islam [2].

The padlocks have various sizes, ranging from very small that often used for locking bags of luggage up to very large sizes to lock fences or warehouses. Although the smart locking system are being developed [3,4,5,6,7,8,9], the classical padlock systems are still widely applied based on the reasons above. Interestingly, although the padlocks are very classic items–invented around the Roman age [1] –its concept meets considerations in new science to model the processes or mechanisms in biological systems, for example the method of pedlock probes. This method has been applied to detect sets of gene sequences with high specificity and excellent selectivity for sequence variants [10], detection and subtyping of seasonal influenza [11], localized detection of specific nucleic acids [12], understanding the molecular docking [13], molecular recognition [14], reaction between enzymes and substrates [15], differentiating single-nucleotide [16], etc.

At present, the need for padlock increases due to massive infrastructure development and it has triggered mass production of padlocks. This implies the structure of keys and locking parts inside the padlocks become simpler. This is why, some people still worry about using the padlock for high security purposes [17].Most of the current padlock architectures are merely rectangular shaped rods which are then grounded in several positions. We call now the grinded parts as holes and the parts between two holes as hills. The number of holes in a key of commercial padlocks is commonly a maximum of four. The number of holes, the size of holes, and the size of hills are specific for a key. Because the number of holes, the size of holes, and the size of the hills are limited, then the possibility of duplicating the same key –if the number of produced padlocks is very large– is possible. We have already experienced that.

As shown in **Figure 1** (a), we have two different padlocks from different brands. Surprisingly, the key in **Figure 1** (a) can open both padlocks. **Figure 1** (b) is unlocking the first padlock and **Figure 1** (c) is unlocking the second padlock using the same key.**Figure 1**(d) is a workshoper duplicating a key using a grinding machine. We only need to pay 1 dollar for his service. The interesting question is, how many different numbers of padlocks can be produced? What parameters affect the number of different padlocks that can be produced?

This seems to be an interesting topic and can be discussed in undergraduate level. The problem begins with an accidental observation of a phenomenon and then thinking and deriving of the formulation to explain such a phenomenon. The mathematical equations should not be too complicated, but an imagination is required to derive the equations. We have done similar works previously, such as initiated by the observation of firework's rod bending [18], discharge of water when wringing wet clothes [19] and segregation of rice when winnowing the tray [20].

The purpose of this paper is to derive the equations to predict the number of different padlocks that can be produced using key structures as shown in **Figure 1** (a). We only focus on keys having identical holes and hills. Indeed, there are some keys having holes with different depths, but we ignore this situation. We also only focus on keys having a maximum of four holes. This limitation is very realistic because most padlocks with the above architecture have keys with up to four holes. For example, the key in **Figure 1** (a) is considered to have three holes. Two bottom holes having different depths are treated as one hole, with the width of which is the sum on widths of the original holes. To best our knowledge, the discussion of physical model for describing the padlock differentials has never been reported.

## 2. Modeling

First, let us consider the case when only one hole was crafted on the key. The hole has a minimum width according to key mechanical structure as well as grinding equipment to craft the hole. Usually, household padlocks are not very precise items. Assume the minimum shift of a grinding cutting wheel to make a hole is *a*. This value may be less than 1 mm. When the hole is made, the grinding cutting wheel can shift at least as far as *a*. If the length of the key portion that penetrating into the padlock body is *L*, the maximum number of shifts of the grinding cutting wheel is $N = L/a$. Suppose the minimum size of a hole and hill that can be made are $h_1$ and $h_2$, respectively. The number of shifts related to those minimum sizes are $n_0 = h_1 / a$ and $m_0 = h_2 / a$, respectively, with $n_0$ and $m_0$ are integers. Next, we only discuss the problem in term of these integers, instead of in term of meter or centimeter. It is clear that $n_0, m_0 \leq N$.

Take a look at **Figure 2** (top). One hole of width $n_0$ is made at the far left end of the key. The width of the hill on the right side is $N - n_0$. This is one different key that can be made. Another different key is possible made if the hole is moved to the right side of *a*, which is identical to a shift of one unit. The number of shifts until the hole reaches the far right position (**Figure 2** (bottom)) is $N - n_0 + 1$.

Next different key is made by crafting hole of one step wider, i.e. $n_0 + 1$. In this condition, the width of the hill on the right side is $N - n_0 - 1$. Different keys can be generated by shifting the hole one unit to the right. The number of shifts until the hole reaches the far right position is $N - n_0$. This is the number of different keys that can be produced if the size of the hole is $n_0 + 1$.

The next different keys are made by crafting a wider hole, i.e., of size $n_0 + 2$. The number of shifts belong to this size is $N - n_0 - 1$. We continue this procedure until the hole has the width of $N$, which has only one choice (no shift). Therefore, the total number of different keys produced by only making one hole is

$$W(1:N,n_0) = 1 + 2 + \ldots + (N - n_0 + 1)$$

$$= \frac{1}{2}(N - n_0 + 1)(N - n_0 + 2) \tag{1}$$

Now we examine the next case where there are two holes and one hill between these holes. The smallest size of a hill that can be made is $m_0$. We start by looking at the condition of all of the holes and hills are located at the left end of the key and all of them have the minimum sizes as illustrated in **Figure 3** (top). Now look at the hole is positioned on the right side. The size of the hill on the right side of this hole is $N - 2n_0 - m_0 = N - (m_0 + n_0) - n_0$. This is one different key. Other different key is obtained by sifting the right hole to the right for every step. The number of shifts until this hole reaches the far right position (formation like in **Figure 3** (bottom).) is $N - (m_0 + n_0) - n_0 + 1$. Furthermore, different keys can be made by widening the rightmost hole by one unit so it become $n_0+1$. This widening results in the width of hill on the right side becomes $N-(m_0+n_0)-n_0-1$, so the number of ways of moving the right most hole becomes $N-(m_0+n_0)-n_0$. We continue this procedure until the rightmost hole occupies all the space on the right side with a width of $N-(m_0+n_0)$ and provides no shifting. As a result, the number of ways to make a key with only changing the position and size of the rightmost hole, and maintaining the position and size of the left hole and the hill in between is

$$s_1 = 1+2+...+\left[N-(m_0+n_0)-n_0+1\right]$$

$$= \frac{1}{2}\left[(N-n_0)-(m_0+n_0)+1\right]\left[(N-n_0)-(m_0+n_0)+2\right]$$

$$= W(1:N-(m_0+n_0),n_0) \qquad (2)$$

The next different key is made by shifting the position of the hill between two holes by one unit to the right. The shift will cause the space for shifting the rightmost hole decreases by one unit. Thus, the number of ways of shifting the position and resizing of the rightmost hole can be obtained from equation (2) by reducing the last term by one unit, i.e.

$$s_2 = 1+2+...+\left[N-(m_0+n_0)-n_0\right]$$

$$= \frac{1}{2}\left[(N-n_0)-(m_0+1+n_0)+1\right]\left[(N-n_0)-(m_0+1+n_0)+2\right]$$

$$= W(1:N-(m_0+1+n_0),n_0) \qquad (3)$$

Next different key is made by shifting the hill between the two holes two steps to the right. The number of ways to change the position and size of the rightmost hole changes to

$$s_3 = W(1:N-(m_0+2+n_0),n_0) \qquad (4)$$

We continue this procedure until the rightmost hole only occupies the far right end with the width of only $n_0$ and the number of compilation is only 1. The total number of hill shifts is $N-2n_0$. Thus, the number of arrangements by changing the position and width of the rightmost hole and the hill between two holes is

$$\Psi(n_0, m_0, 0) = \sum_j s_j$$

$$= \sum_{j=0}^{N-2n_0} W(1: N-(n_0+m_0+j), n_0) \tag{5}$$

We can easily prove that if $j = N - 2n_0$ then $W(1: N-(N-2n_0+n_0), n_0)$ $W(1: N-(N-n_0), n_0) = W(1: n_0, n_0) = 1$.

Next, we begin to move the holes in the left position one step each. If this hole is moved one step to the right, the space to make variations of hole and hill to the right is as if reduced by one compared to the Initial condition. Thus, the number of variation satisfies equation (5) but with replacing $N$ with $N$-1, resulting

$$\Psi(n_0, m_0, 1) = \sum_{j=0}^{N-2n_0-m_0} W(1:(N-1)-(m_0+j+n_0), n_0) \tag{6}$$

If the left hole is moved two sites to the right, the number of ways is

$$\Psi(n_0, m_0, 2) = \sum_{j=0}^{N-2n_0-m_0} W(1:(N-2)-(m_0+j+n_0), n_0) \tag{7}$$

And so on so that we get the general equation for the shift as far as $k$, i.e.,

$$\Psi(n_0, m_0, k) = \sum_{j=0}^{N-2n_0-m_0} W(1:(N-k)-(m_0+j+n_0), n_0) \tag{8}$$

The maximum number of shifts is $N - m_0 - 2n_0$. Thus, the number of ways through shifting only the leftmost hole is

$$\Psi(n_0, m_0) = \sum_{k=0}^{N-m_0-2n_0} \Psi(n_0, m_0, k) \tag{9}$$

Next different key is obtained by changing the width of the left hole. When the width is enlarged to $n_0 + 1$ then the number of shifts becomes $N - m_0 - 2n_0 - 1$ and the number of ways the arrangement becomes

$$\Psi(n_0+1, m_0) = \sum_{k=0}^{N-m_0-2n_0-1} \Psi(n_0, m_0, k) \tag{10}$$

When the width of the leftmost hole is changed to $n_0+2$, the number of shifts becomes $N - m_0 - 2n_0 - 2$ and the number of arrangements becomes

$$\Psi(n_0+2, m_0) = \sum_{k=0}^{N-m_0-2n_0-2} \Psi(n_0, m_0, k) \tag{11}$$

We continue the process until the maximum width for the leftmost hole is reached. The allowed maximum width of the leftmost hole is $N - m_0 - n_0$. This means the maximum additional width that can be generated is $N - m_0 - 2n_0$. The number of ways of widening this hole is

$$\Psi(n_0 + (N - m_0 - 2n_0), m_0) = \sum_{k=0}^{N-m_0-2n_0-(N-m_0-2n_0)} \Psi(n_0, m_0, k) = \sum_{k=0}^{0} \Psi(n_0, m_0, k) = 1 \tag{12}$$

And then we finally obtain that the total number of different keys that can be produced when two holes are made is

$$W(2:N, n_0, m_0) = \sum_{i=0}^{N-m_0-2n_0} \Psi(n_0 + i, m_0)$$

$$= \sum_{i=0}^{N-m_0-2n_0} \left( \sum_{k=0}^{N-m_0-2n_0-i} \Psi(n_0, m_0, k) \right)$$

$$= \sum_{i=0}^{N-m_0-2n_0} \left( \sum_{k=0}^{N-m_0-2n_0-i} \sum_{j=0}^{N-2n_0-m_0} W(1:(N-k)-(m_0+j+n_0), n_0) \right) \tag{13}$$

Next we discuss the keys containing three holes. The base position of the holes and hills is shown in **Figure 4**. There are three holes, each having a width of $n_0$ where two adjacent holes are separated by a hill of width $m_0$.

Let us first consider the case when the hole and the hill in the far left position do not change in both position and size. Two holes and one hill on the right claim a space $N' = N - (n_0 + m_0)$ width. The number of variations of two holes and one hill on the right satisfies equation (13) with only replacing $N$ with $N'$. Therefore, we have

$$W(2:N', n_0, m_0) = W(2:N-(m_0+n_0), n_0, m_0) \tag{14}$$

Next, we move the leftmost hill to the right by one step. Two holes and one hill on the right occupy a space of $N'-1$ width. Thus, the number of variations if the second hill is moved to the right by one step is

$$W(2:N'-1, n_0, m_0) = W(2:N-(m_0+n_0)-1, n_0, m_0) \tag{15}$$

If the second hill is moved to the right by two steps, the number of variations is

$$W(2:N'-2, n_0, m_0) = W(2:N-(m_0+n_0)-2, n_0, m_0) \tag{16}$$

The number of shifting the left hill to the right is $[N-(m_0+n_0)]-(n_0+m_0+n_0) = N-2m_0-3n_0$. Thus, the total number of ways if the leftmost hole remains unchanged but with position and size of others are changed becomes

$$\Theta = \sum_{\ell=0}^{N-2m_0-3n_0} W(2:N-(m_0+n_0)-\ell, n_0, m_0) \tag{17}$$

Next, we change the position of the leftmost hole by shifting it step by step to the right. The initial shift is identical to decreasing $N$ by 1 to become $N$-1 at the upper limit of the summation in equations (17). The rightmost position of the left hole is obtained after the maximum shift is $N-2m_0-2n_0$. Thus, the number of shifts to reach this maximum shifting is $N-2m_0-2n_0-n_0 = N-2m_0-3n_0$. Finally, the total number of ways of producing keys containing three holes is

$$W(3:N, m_0, n_0) = \sum_{p=0}^{N-2m_0-3n_0} \sum_{\ell=0}^{(N-p)-2m_0-3n_0} W(2:N-(m_0+n_0)-\ell, n_0, m_0) \tag{18}$$

Finally, we discuss the conditions when the key has four holes as illustrated in **Figure 5**. We begin by forming the basic condition where four holes and three hills are at the left end of the key and each has a minimum size.

If the leftmost hole and hill does not change, then the sum of the variations in the three holes and the two hills on the right satisfies equation (18) with simply replacing $N$ with $N-(n_0+m_0)$. The number of ways is

$$W(3:N-(n_0+m_0),m_0,n_0) \tag{19}$$

If the rightmost hill is moved to the right one step then the number of ways is

$$W(3:N-(n_0+m_0)-1,m_0,n_0) \tag{20}$$

If the rightmost hill is moved to the right two steps, the number of ways is

$$W(3:N-(n_0+m_0)-2,m_0,n_0) \tag{21}$$

The number of shifts to the rightmost hills is $N-3m_0-4n_0$. Thus, the total number of ways is

$$\Phi = \sum_{q=0}^{N-3m_0-4n_0} W(3:N-(n_0+m_0)-q,m_0,n_0) \tag{22}$$

Finally, the position of the leftmost hole is shifted to the right for every step. Each shift seems to reducing the width by 1. The number of shifts to the right is $N-3m_0-4n_0$. Thus, the total number becomes

$$W(4:N,m_0,n_0) = \sum_{r=0}^{N-3m_0-4n_0} \left( \sum_{q=0}^{(N-r)-3m_0-4n_0} W(3:N-(n_0+m_0)-q,m_0,n_0) \right) . \tag{23}$$

We then obtain a recursion formula for arbitrary number of holes

$$W(u+1:N,m_0,n_0) = \sum_{r=0}^{N-um_0-(u+1)n_0} \left( \sum_{q=0}^{(N-r)-um_0-(n+1)n_0} W(u:N-(n_0+m_0)-q,m_0,n_0) \right) \tag{24}$$

with $u$ is the number of holes, as long as $N-um_0-(u+1)n_0 \geq 0$ and $u \geq 2$.

All of the summations above are finite terms so that the final solution can always be obtained. Using a free accessed symbolic mathematics, we obtained [21]

$$W(2:N,m_0,n_0) = \frac{1}{24}(S+1)^2(S^3+7S^2+22S+24) \tag{25}$$

$$W(3:N,m_0,n_0) = \frac{(Q+1)}{5040}(30Q^6 + 425Q^5 + 2662Q^4$$

$$+ 8923Q^3 + 16172Q^2 + 14628Q + 5040) \tag{26}$$

$$W(4:N,m_0,n_0) = \frac{U+1}{120960}(80U^8 + 1735U^7 + 16709U^6 + 91273U^5 + 303023U^4$$

$$+ 615832U^3 + 738084U^2 + 471024U + 120960) \tag{27}$$

with

$$S = N - 2n_0 - m_0, \tag{28}$$

$$Q = N - 3n_0 - 2m_0, \tag{29}$$

and

$$U = N - 4n_0 - 3m_0. \tag{30}$$

## 3. Results and Discussion

Now let us do some simulations. The portion of key inserted into the padlock is usually not too long. Generally, the length of this portion is around 1.5 cm to 2.0 cm. If we assume that the length of this portion is 2.0 cm and the smallest portion that can be grinded has a length of 1 mm, the length of the key corresponds to $N = 20$ sites. The true smallest value depends on the machine used to make the key and the padlock. However, for most of household padlocks, the number does not vary too far from $N = 20$.

**Figure 6** is the number of different keys that can be made with a number of holes between 1 and 4 at various hole and hill lengths. In this case, we assume that the hole length and the hill length are identical. The total number of keys is defined as

$$W(tot) = W(1:N,n_0) + W(2:N,m_0,n_0) + W(3:N,m_0,n_0) + W(4:N,m_0,n_0) \tag{31}$$

It appears that as the minimum size of hole and hill increases, the number of different keys decreases. This is due to the number of ways of shifting and widening the hole and the hill decrease. For $N$ = 15, 20, and 25, we obtain a scaling relationship

$$W(tot) \propto e^{-\gamma m_0} \tag{32}$$

with $\gamma$ = 6.16, 6.82, and 7.10 for $N$ = 15, 20, and 25, respectively. The scaling factor rises slightly with increasing the key length. We can show that $\gamma \propto N^{1/3}$ as shown in inset of **Figure 6**.

**Figure 7** shows the number of different keys as a function of $n_0$ at different $m_0$. We maintain $N$ = 20. The number of different keys decreases as $n_0$ increases. The scaling relationship according to equation (32) is also observed here, but with a nearly constant scaling factor. If $m_0$ increases, the number of different keys that can be made decreases.

We also estimate the number of different keys – in combination formulation – if the width of hole and hill are the same. If the widths of hole and hill are identical, $m_0 = n_0 = 1$, the number of different keys for $N$ = 20 is the same as combination of two states, with each has a width of unity, having the values ranging from 0 to 20 with the sum of both number is 20. In general, if the total number of holes and hills is $M$, the number of holes is $n$ and the number of hills is $M-n$, then the number of combinations up to a maximum of four holes is

$$W(tot) = \sum_{n=0}^{4} \frac{M!}{n!(M-n)!} \tag{33}$$

It should be noted here that $M \neq N$ but $M = N/n_0 = N/m_0$.

**Figure 8** are the calculation results using equation (33) at different $M$: $M$= 15, 20, and 24. Calculations were made for selected $n_0 = m_0$ that produces $M$ as an integer. For example, at $N$ = 15, the calculation is only performed for $n_0 = 1$, $n_0 = 3$, and $n_0 = 5$.

We get the exponential equation (32) with γ changes slightly with increasing $N$. **Figures 6** and **8** show that the number of different key calculated using equation (31) changes exponentially, resemble those of calculated using Eq. (33). Both are different in scaling parameter because in equation (34) the width of the hole or hill is always a multiple of $m_0$ or $n_0$, whereas in equations (31) $m_0$ and $n_0$ are the minimum widths of holes and hills. In Eq. (31), the width of holes does not need to be a multiple of $n_0$ and the width of hills does not need to be a multiple of $m_0$. This is why the results of calculations using equation (31) are larger than those of using equation (33).

Reducing $n_0$ or $m_0$ has the meaning of making keys with higher accuracy. It can be seen from **Figure 6** or **8** that the more accurate the key is, the more variations can be made. We also found that for keys with a length of 2 cm and a minimum size of holes and hills about 3 mm, the number of different keys is only around a half of million. This amount becomes less if the portion of the key that enters the padlock is only about 1.5 cm (the number commonly found for household locks).

Let us consider the limiting cases when $N \gg n_0$, $m_0$ such that one can approximate $S \approx Q \approx U \approx N$. Under this condition, we obtain from Eqs. (25)-(27) that $W(2:N,m_0,n_0) \propto N^5$, $W(3:N,m_0,n_0) \propto N^7$, and $W(4:N,m_0,n_0) \propto N^9$. This behavior can be obtained from the recursive Eq. (24) as follow. If $N \gg n_0$, $m_0$ one may approximate

$$W(u:N-(n_0+m_0)-q,m_0,n_0) \approx W(u:N,m_0,n_0)$$
(34)

so that

$$W(u+1:N,m_0,n_0) \approx \sum_{r=0}^{N}\left(\sum_{q=0}^{N} W(u:N,m_0,n_0)\right)$$

$$\approx (N+1)^2 W(u:N,m_0,n_0) \approx N^2 W(u:N,m_0,n_0) \tag{35}$$

Since $W(2:N,m_0,n_0) \propto N^3$, we then obtain $W(u:N,m_0,n_0) \propto N^{2u+1}$ for $u > 1$ which is consistent with Eqs. (25)-(27).

As final notes we obtained that the number of different keys for the above key architecture is very limited. For example, for $N = 15$ and $n_0 = m_0 = 3$, the different keys that can be made (based on Fig. 6) is approximately $\exp(8) \approx 3{,}000$ and for $N = 15$ and $n_0 = m_0 = 2$ the different keys that can be made is approximately $\exp(10) \approx 22{,}000$. This variation is not soo high so that duplication of the a certain key may happen.

## 4. Conclusion

We have derived equations for estimating the number of different keys that can be made based on key architecture as in Fig. 1(a). We obtained a general recursive formula for estimating the number of different keys as a function of the number of holes. For very precise keys, we obtained the number of different keys increase as $N^{2u+1}$ with $N$ is the key length and $u$ is the number of holes. We argue the equation might be applied to estimating the number of configurations in the padlock probes method of biological systems.

## References

[1] Haddad N A 2016 *Mediterran. Archaeol. Archaeometry* **16** 53
[2] S. S. Çelik SS 2015 *IntJSCS* **3** 96
[3] Adiono T, Fuada S, Anindya S F, Purwanda I G and Fathany M Y 2019 *Int. J. Adv. Comp. Sci. Appl.* **10** 445
[4] Kumbhar N N and Deshmukh P V M 2017 *Int. J. Res. Appl. Sci. Eng. Technol.* **5** 507
[5] R. Satoskar and Mishra A 2018 *Int. J. Comp. Sci. Inf. Technol.* **9** 132

[6] Li W, Li H, Gong A, Ou Y and Li M 2018 *J. Phys.: Conf. Ser.* **1069** 012134

[7] Shashank K, Gopalakrishna M T and Hanumantharaju M C *2014 Proc. 3rd International Conference on Frontiers of Intelligent Computing: Theory and Applications (FICTA)* 685

[8] Hadis M S, Palantei E, Ilham A A and Hendra A 2018 *International Conference on Information and Communications Technology (ICOIACT)*, doi:10.1109/ICOIACT.2018.8350767

[9] Sunehra D 2019 *IOSR J. Eng.* **09** 29

[10] Banér J, Nilsson M, Mendel-Hartvig M and Landegren U 1998 *Nucleic Acids Res.* **26** 5073

[11] Neumann F, Hernandez-Neuta I, Grabbe M, Madaboosi N, Albert J and Nilsson M 2018 *Clinical Chem.* **64** 1704

[12] Nilsson M, Malmgren H, Samiotaki M, Kwiatkowski M, Chowdhary B P and Landegren U 1994 *Science* **265** 2085

[13] Tripathi A and Bankaitis V A 2017 *J. Mol. Med. Clin. Appl.* **2**, 1

[14] Ben-Naim A 2018 arXiv:1806.03499

[15] Schneider H –J 2015 *Int. J. Mol. Sci.* **16** 6694

[16] Nilsson M, Krejci K, Koch J, Kwiatkowski M, Gustavsson P and Landegren U 1997 *Nature Genetics* **16** 252

[17] Lee F W W and Ibrahim A B 2018 *J. Inf. Sys. Technol. Manag.* **3** 26

[18] Abdullah M, Khairunnisa S and Akbar F 2014 *Eur. J. Phys.* **35** 035019

[19] Rahmayanti H D, Utami F D and Abdullah M 2016 *Eur. J. Phys.* **37** 065806

[20] Munir R, Rahmayanti H D, Murniati R, Rahman D Y, Utami F D, Viridi S and Abdullah M 2020 *Granular Matter* **22** 24.

[21] We used Wolframalpha to do the calculations explicitly (accessed on January 2, 2020)

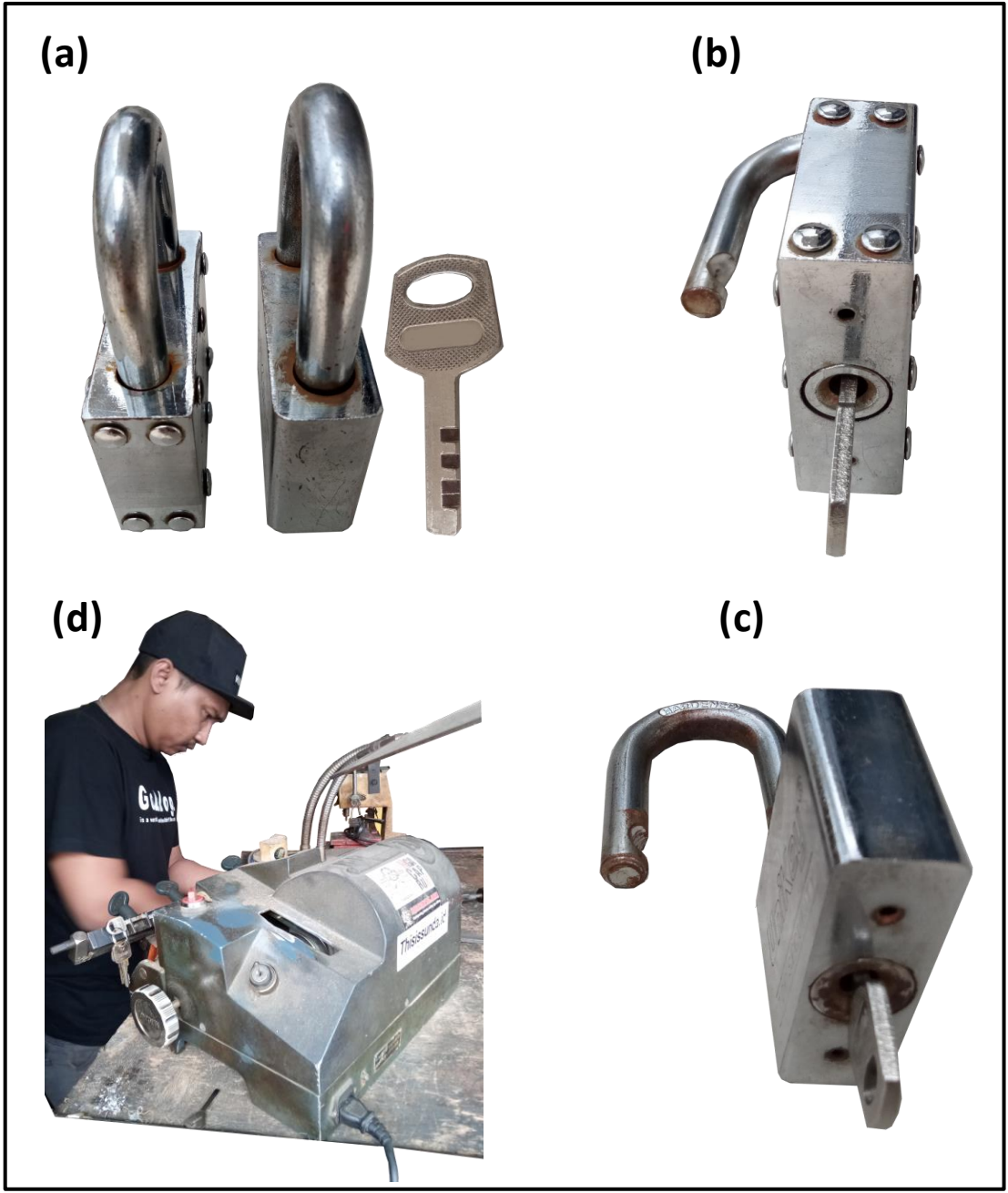

**Figure 1** (a) Two padlocks and a key that can open both padlocks. (b) When the key is opening the first padlock. (c) When the same key is opening the second padlock. (a) A worker is duplicating a key using a grinding machine.

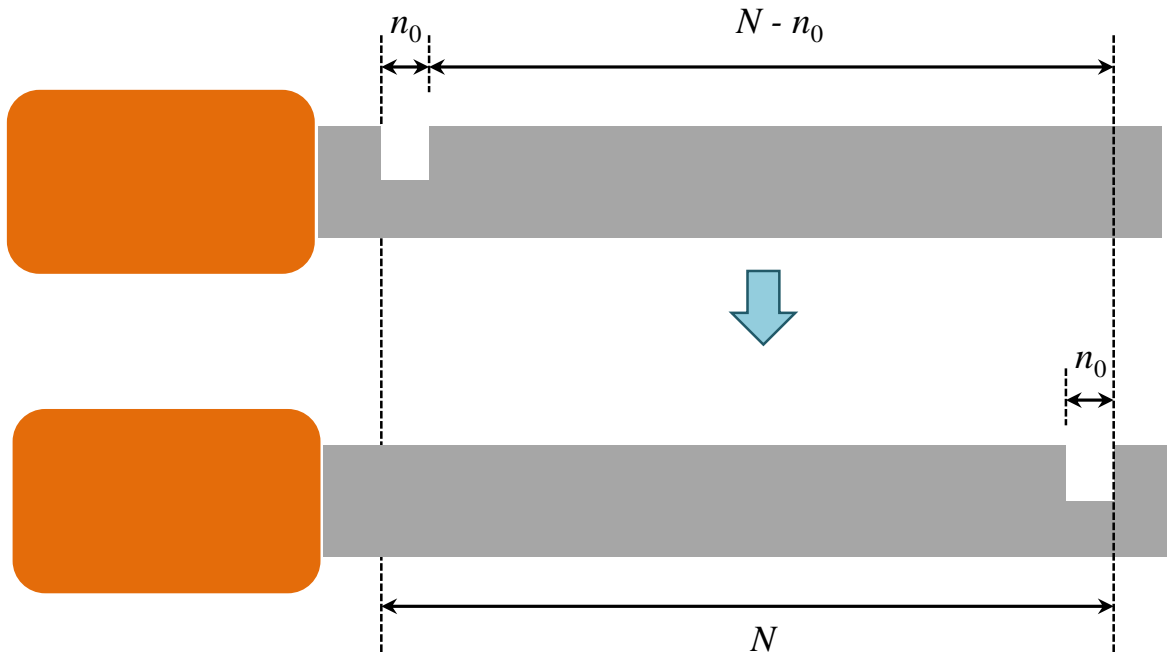

**Figure 2** A key contains only one hole. We can change the position and the width of the hole to produce different keys.

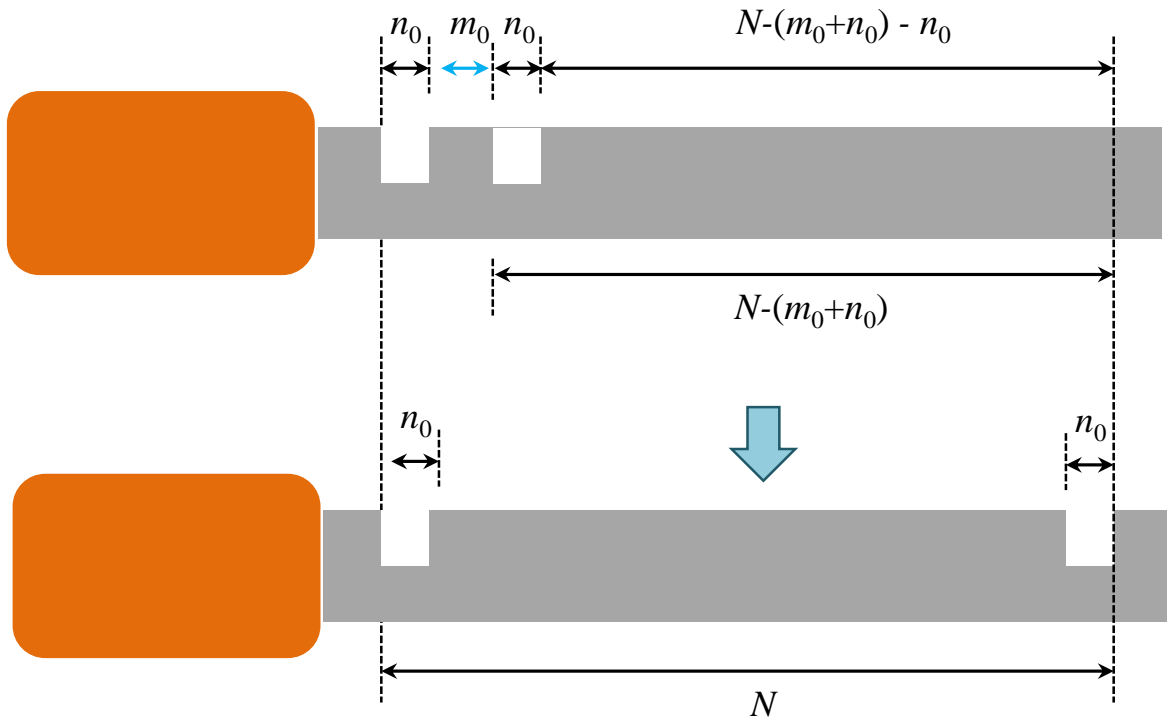

**Figure 3** A key contains two holes. We can change the position and width of holes and hills to produce different keys.

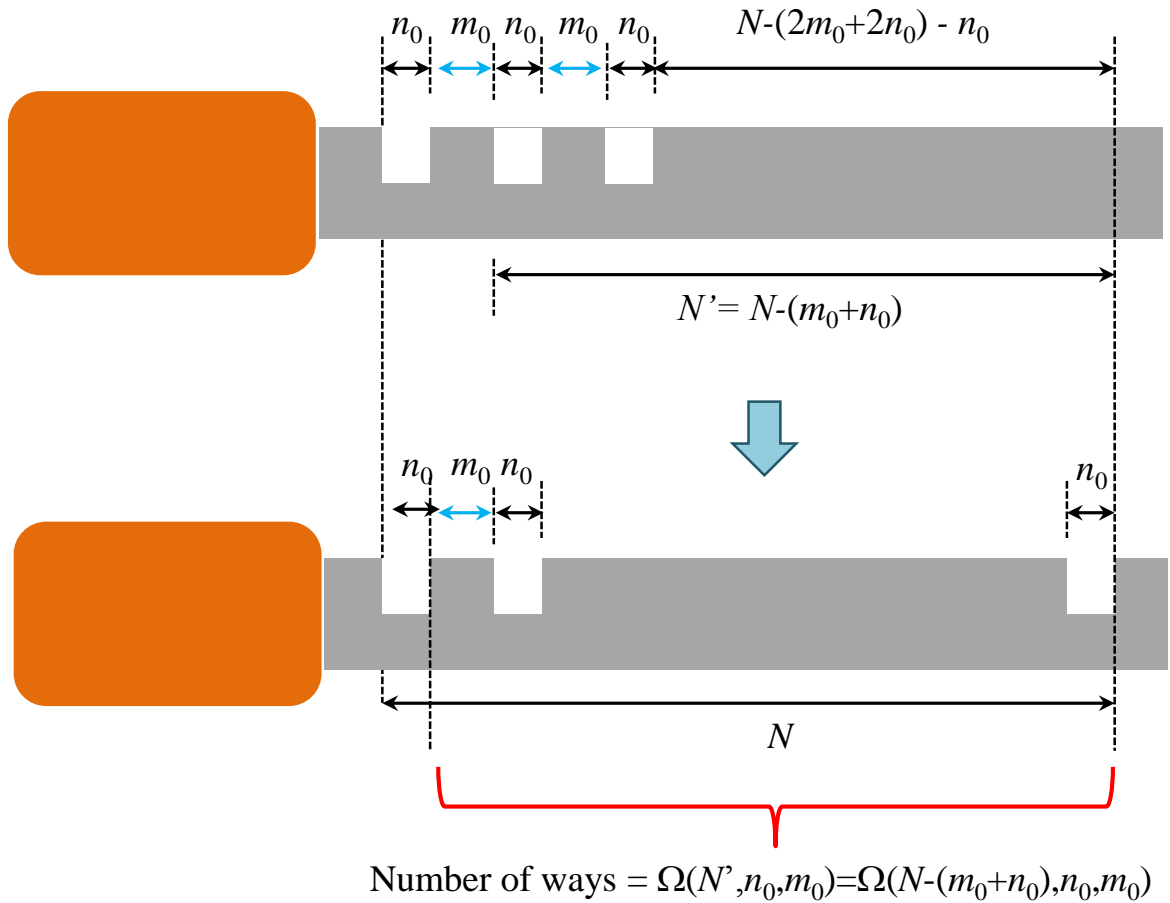

**Figure 4** A key contains three holes. We can change the position and width of holes and hills to produce different keys.

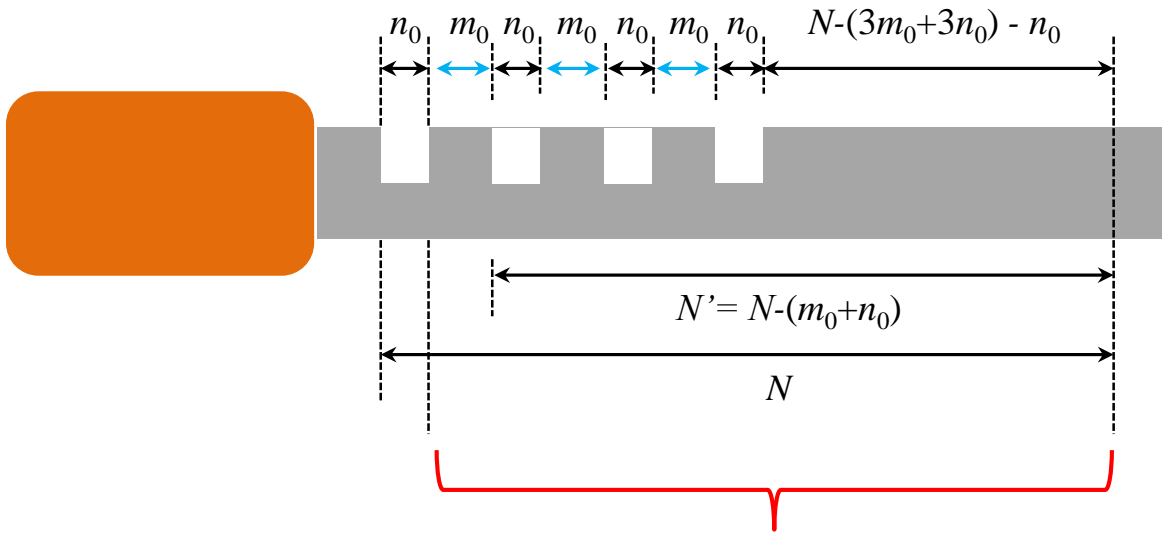

Number of ways = $\Theta(3:N',n_0,m_0)=\Theta(3:N-(m_0+n_0),n_0,m_0)$

**Figure 5** A key contains four holes. We can change the position and width of holes and hills to produce different keys.

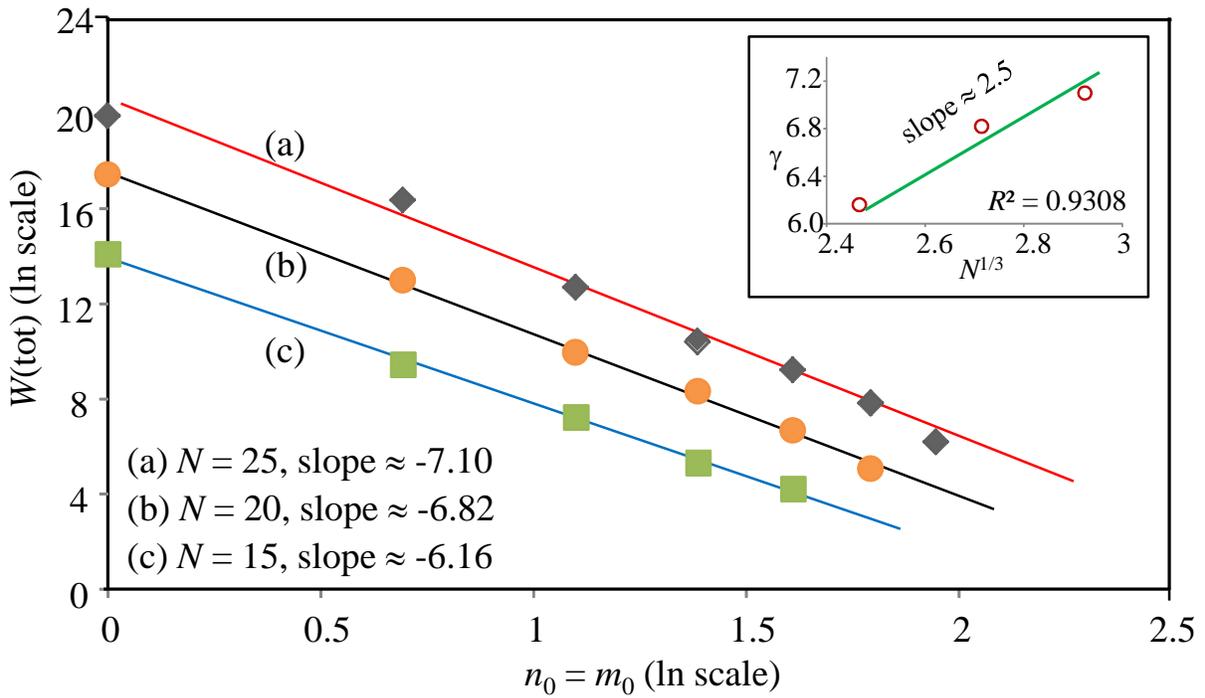

**Figure 6** Number of total different keys that can be produced by making up to four holes as a function of the width of the hole and hill. The width of the hole and the width is assumed to be identical. Calculations were performed for three different lengths of the keys: (a) $N = 25$ steps, (b) 20 steps, and (c) 15 steps. Inset is the curve of scaling factor as a function of 1/3 power of key length.

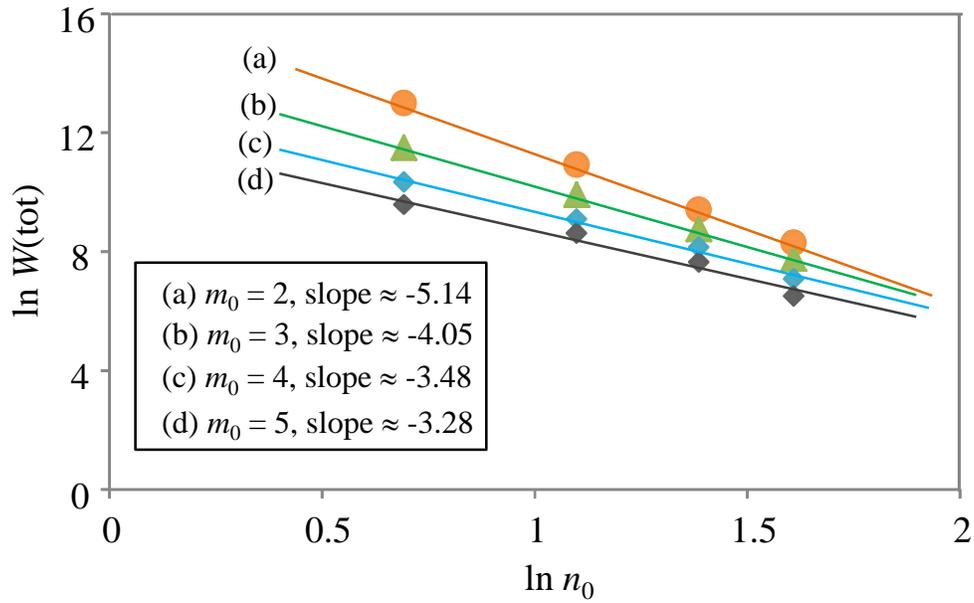

**Figure 7** Number of total different keys that can be produced by making up to four holes as a function of the width of the hole at different width of the hole. We calculated for different values of the hill: (a) $m_0 = 2$, (b) $m_0 = 3$, (c) $m_0 = 4$, and (d) $m_0 = 5$ and the total length of the keys was kept as $N = 20$ steps.

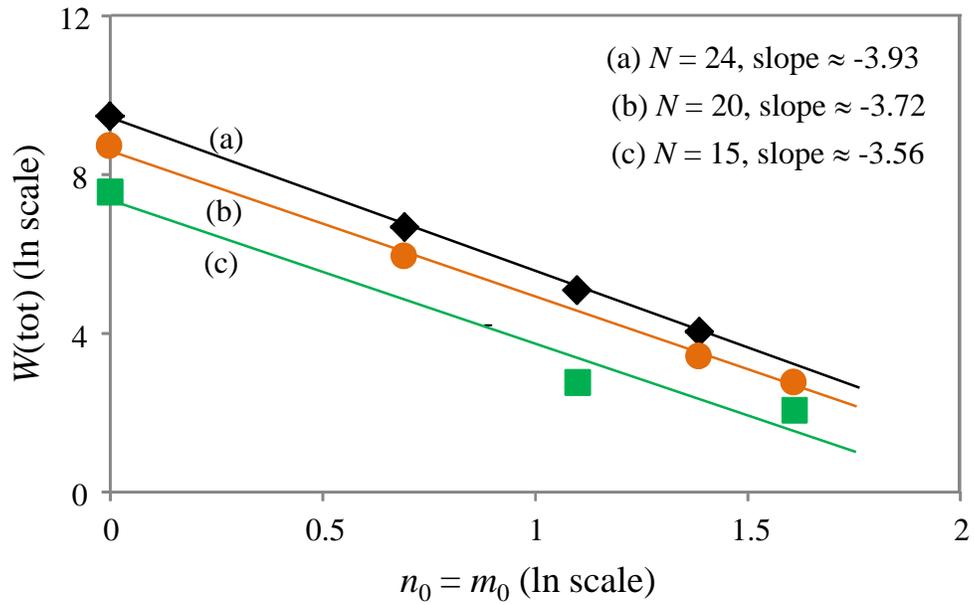

**Figure 8** Number of total different keys that can be produced by making up to four holes as a function of the width of the hole and hill calculated using Eq. (33). The width of the hole and the width is assumed to be identical. Calculations were performed for three different lengths of the keys: (a) $N = 24$ steps, (b) 20 steps, and (c) 15 steps.